\documentclass[twocolumn]{aastex631}

\usepackage{graphicx}
\usepackage{amsmath}
\usepackage{cleveref}
\usepackage{float}
\usepackage{pgffor}
\usepackage{tabularx}
\usepackage{array}
\usepackage{xcolor}
\newcolumntype{Y}{>{\raggedright\arraybackslash}X}

\usepackage{makecell}

\graphicspath{{./}}

\begin{document}

\title{Uber-calibration-based Two-level Radial-velocity Zero-point Correction for APOGEE}

\shorttitle{Two-level RV Zero-point Correction for APOGEE}
\shortauthors{Zhang \& Yuan}

\author[0009-0005-7743-6229]{Jinming Zhang}
\affiliation{Institute for Frontiers in Astronomy and Astrophysics, Beijing Normal University, Beijing, 102206, China}
\affiliation{School of Physics and Astronomy, Beijing Normal University, Beijing, 100875, China}

\author[0000-0003-2471-2363]{Haibo Yuan}
\affiliation{Institute for Frontiers in Astronomy and Astrophysics, Beijing Normal University, Beijing, 102206, China}
\affiliation{School of Physics and Astronomy, Beijing Normal University, Beijing, 100875, China}

\author[0000-0003-0220-7112]{Zhijia Tian}
\affiliation{Department of Astronomy, Key Laboratory of Astroparticle Physics of Yunnan Province, Yunnan University, Kunming 650200, China}

\correspondingauthor{Haibo Yuan}
\email{yuanhb@bnu.edu.cn}

\begin{abstract}
APOGEE delivers high-resolution, near-infrared spectroscopy and highly precise radial velocities (RVs), with extensive repeat observations spanning more than a decade. Long-term RV repeatability is nevertheless limited by instrumental zero-point systematics: exposure-to-exposure drift and fiber-dependent offsets that are only partially tracked by lamp-based wavelength calibration, and that can evolve as fibers are serviced or replaced. We infer and remove these zero points directly from the published per-visit RVs using an uber-calibration framework based on shared repeat observations. We solve sequentially for two sets of minimal calibration units: (plate, night) exposure-level offsets and time-dependent (fiber, observing year) offsets. We apply the method independently to APO and LCO DR17 and to APO DR19. After correction, the same-fiber repeatability within two days reaches a limiting precision of $\sim 14~\mathrm{m\,s^{-1}}$, and the long-baseline single-visit precision improves by a factor of two from 98.2 to 44.7~$\mathrm{m\,s^{-1}}$. We release per-visit, value-added RV catalogs for DR17 (APO+LCO) and DR19 (APO) containing 3,501,727 visits (3,336,141 fully corrected) of 930,519 unique stars, enabling more reliable RV variability, binary, and Galactic-dynamics studies.
\end{abstract}


\section{Introduction}

Radial velocity is a cornerstone observable in stellar astronomy: it enables the identification of binary and multiple systems, the detection of exoplanets, and the diagnosis of pulsation and atmospheric activity, and it is a key input to studies of cluster dynamics and Galactic archaeology \citep{tian2020catalog,kovalev2024,guo2025sb2orbits}. Over the past two decades, multi-object fiber spectroscopic surveys have expanded radial-velocity work from star-by-star programs to catalogs containing millions of measurements; RAVE, SEGUE, APOGEE, LAMOST, \textit{Gaia} RVS, and DESI are prominent examples \citep{rave2006,yanny2009segue,majewski2017apogee,cui2012lamost,gaia2016gaia,desidr1}. The uncertainty of a single measurement is commonly separated into a random component, dominated by the spectral signal-to-noise ratio, and a systematic component, driven by wavelength calibration and instrumental zero-point drift. Random errors can be reduced by higher signal-to-noise ratios or additional epochs, but systematic errors do not average down with sample size and must be removed through calibration. At high signal-to-noise ratios, systematics therefore set the effective precision floor, and for surveys spanning a decade they can also vary with time.

Among these surveys, APOGEE is distinctive: it delivers high-resolution, near-infrared spectroscopy from both hemispheres and has repeatedly observed many of the same targets for more than a decade, combining large sample size with long time baselines. It is therefore both a workhorse for Galactic archaeology and a powerful testbed for radial-velocity systematics \citep{majewski2017apogee,wilson2019}. As in other fiber surveys, APOGEE's precision is ultimately limited by the stability of the wavelength solution, and SDSS pipeline and instrument developments have progressively pushed this calibration to finer time sampling and denser reference line sets. In SDSS-III/IV, the wavelength solution was derived from hollow-cathode lamp exposures on a per-period basis (coarse time granularity) \citep{nidever2015}. In SDSS-V, the reduction pipeline solves for the wavelength solution daily \citep{apogeedr19}, and a Fabry--P\'erot etalon was added to provide a dense, uniform set of calibration lines; together with back-pressure regulation and octagonal-core fibers, these upgrades target improved radial-velocity precision \citep{wilson2022}.

These improvements target a common failure mode: time-dependent drift in the pixel-to-wavelength mapping. However, not all of the radial-velocity zero point is captured by the wavelength calibration. First, because the wavelength solution is not re-derived for every exposure, and the instrument state evolves over the interval covered by a single solution, each exposure can carry its own overall velocity offset. Second, each fiber traces a distinct optical path, and the calibration lamp illuminates the fibers uniformly via a screen, whereas starlight enters as a guided spot whose illumination pattern depends on guiding and atmospheric conditions; this mismatch introduces fiber-to-fiber zero-point differences that are absent from the calibration exposures and therefore cannot be removed by the lamps alone. The second effect has already been mitigated with substantial success: \citet{saydjari2025} modeled stellar and sky components jointly and explicitly accounted for fiber zero-point differences, improving the plate-era precision to about 30~$\mathrm{m\,s^{-1}}$. Beyond this, time provides an additional dimension: fiber zero points are not constant, because re-termination and replacement of fibers, routine maintenance, and changes in the illumination feed slowly alter fiber states and hence their zero points. Modeling this evolution offers a further route to reducing the remaining systematic error.

In this work we take a complementary approach. Rather than altering the spectral extraction, we infer and remove instrumental zero points directly from the published radial velocities. Our model incorporates both instrumental and temporal structure: we distinguish exposure-to-exposure and fiber-to-fiber offsets, and we allow fiber zero points to evolve with time. The calibration strategy traces back to the ``uber-calibration'' developed for photometric surveys \citep{ubercali2008} and follows the framework of \citet{zhang2026lrs,zhang2026b} for spectroscopic radial-velocity calibration: each independently drifting observing stage is treated as a minimal calibration unit, and the zero points of all units are solved for simultaneously using their shared repeat observations. We apply this framework to two APOGEE data releases and to both telescopes, and we deliver a per-visit value-added radial-velocity catalog designed for long-baseline repeatability studies, the identification of radial-velocity variables, and Galactic kinematics.

This paper is organized as follows. Section~2 describes the data and the construction of the calibration sample. Section~3 introduces the two-level zero-point model and defines the calibration units. Section~4 presents the resulting zero-point solutions. Section~5 validates them using repeat observations and stars observed at both sites. Section~6 details the released catalog. Section~7 summarizes our conclusions.

\section{Data}

\subsection{The APOGEE Survey and Spectrographs}

APOGEE is a Galactic stellar survey carried out with two near-infrared, high-resolution spectrographs (APOGEE-N and APOGEE-S; \citealt{majewski2017apogee}). They are mounted on the 2.5~m telescope at Apache Point Observatory (APO) and the du Pont 2.5~m telescope at Las Campanas Observatory (LCO), respectively; each is fed by 300 fibers and covers 15000--17000~\AA\ at $R\approx 22{,}500$ \citep{wilson2019}. The data used in this work are bounded by the SDSS Nineteenth Data Release (DR19): in addition to re-issuing all SDSS-IV-era observations unchanged (i.e., DR17, 2011 August to 2021 January), DR19 includes new SDSS-V observations taken from 2020 October to 2023 July (APO 2.5~m only) \citep{apogeedr19}.

During SDSS-III/IV, the focal-plane fiber configuration was set by plug plates: fiber connectors were inserted into precisely drilled holes in aluminum plates matched to the target positions, and each observation corresponded to one plate. The SDSS observing cadence shows strong seasonal and lunar-phase modulation: new observations predominantly begin during dark time---at LCO, 82.9\% of observations begin between the 7th and 12th day of the lunar month (after new moon and before full moon)---while observations beginning after full moon are substantially rarer. APO is shut down for about two months each summer, so observing years typically run from August--September to May--June of the following year, with most inter-annual gaps falling in August--October. Beginning with SDSS-V (the DR19 era), the plug-plate system was replaced by the robotic fiber positioning system (FPS), and a Fabry--P\'erot etalon calibration lamp (FPI) was introduced as part of a set of external hardware upgrades aimed at improving the radial-velocity precision \citep{wilson2022}.

The DR17 data were reduced with the APOGEE data reduction pipeline (DRP; \citealt{nidever2015}), while the DR19 data were processed with its successor, redux~1.3 \citep{apogeedr19}. The two pipelines differ primarily in their wavelength-calibration schemes: in DR17, the wavelength solution was re-determined from arc-lamp exposures roughly every two weeks \citep{nidever2015}, whereas in DR19 it is solved daily and makes use of the FPI calibration lamp \citep{apogeedr19}.

\subsection{Data Products: the DR17 and DR19 allVisit Catalogs}

Our primary data are the DR17 and DR19 allVisit tables \citep{abdurrouf2022,apogeedr19}, where a ``visit'' is the extracted spectrum of a target from a single observing epoch. The DR17 allVisit catalog contains 2,659,178 visits of 657,135 unique stars; the DR19 catalog (APO 2.5~m only) contains 844,217 visits of 390,838 unique stars.

For each visit, we use the heliocentric-corrected radial velocity (VHELIO in DR17; vrad in DR19). The calibration sample---the visits that define the zero points---is selected using the cuts listed below, quoted first for DR17 (APO+LCO) and then for DR19 (APO only):

\begin{enumerate}
\item Restricting to APO 2.5~m and LCO 2.5~m visits with finite velocities and finite, positive velocity uncertainties leaves 2,580,307 visits in DR17 and 829,653 visits in DR19. A small subset of the DR19 rows is clearly untrustworthy: their velocity fits are poor, with a median $\chi^{2}$ of 9.0 against 2.4 for the rest of the catalogue, yet they carry non-physical velocity uncertainties with a median of $\sim 10^{-23}~\mathrm{km\,s^{-1}}$. Since our zero points are inverse-variance weighted medians of same-star visit pair velocity differences, a single such row outweighs a normal one by many orders of magnitude and pins an entire node pair to that one star. We therefore require the reported DR19 uncertainty to exceed 1~$\mathrm{m\,s^{-1}}$, which removes 2,332 visits.
\item Requiring a median full-spectrum signal-to-noise ratio ${\rm S/N} > 50$ leaves 1,276,501 visits in DR17 and 607,726 in DR19. This cut only determines which visits define the zero points; the resulting corrections are applied to all visits.
\item Requiring ${\tt N\_COMPONENTS} \le 1$ leaves 1,262,742 visits in DR17 and 602,482 in DR19. Here ${\tt N\_COMPONENTS}$ is the number of peaks detected in the cross-correlation function; visits with multiple components are typically binaries or composite sources with unreliable velocities.
\item Requiring a per-star velocity scatter $\sigma(v) < 1$~km~s$^{-1}$ leaves 1,035,354 visits of 425,998 unique stars in DR17 and 537,370 visits of 286,132 unique stars in DR19. This removes velocity-variable stars (binaries, pulsators, etc.), so genuine astrophysical motion is not absorbed into the systematic corrections.
\end{enumerate}

\section{Method}

\subsection{Two-level Zero-point Framework and the $\chi^2$ Model}

We model the systematic errors in the APOGEE radial velocities (RVs) as the sum of two zero-point terms: an exposure-level term $\mu_{\rm e}$ and a fiber-level term $\mu_{\rm f}$. Applying both corrections gives
\begin{equation}
v_{\rm corr} = v - \mu_{\rm e} - \mu_{\rm f}.
\label{eq:vcorr}
\end{equation}

Our calibration follows the uber-calibration framework \citep{ubercali2008,zhang2026b}: each observational stage that drifts independently is treated as a minimal calibration unit, and the zero points of all units are solved simultaneously using repeat observations that link units. In our application, the two unit types are exposures and fibers.

Both levels minimize the same $\chi^2$,
\begin{equation}
\chi^{2} = \sum_{i<j} n_{ij}\,\left[\langle v_i - v_j\rangle - (\mu_i - \mu_j)\right]^{2},
\label{eq:chi2}
\end{equation}
where $\langle v_i - v_j\rangle$ is the weighted median of the visit-to-visit velocity differences formed across units $i$ and $j$, using weights $w = 1/(\sigma_i^{2} + \sigma_j^{2})$. This robust central estimator plays the same role as the Gaussian-fitted central value in \citet{zhang2026b}, providing resilience to a small number of RV-variable outliers; we adopt the weighted median because the $\Delta$RV distribution is sharply peaked with heavy tails, and the median is insensitive to the tails. The factor $n_{ij}$ is the number of cross-unit visit pairs between units $i$ and $j$, and the sum runs over all unit pairs with at least one such pairing. The exposure- and fiber-level problems are isomorphic: we run the same solver twice, changing only the definition of the nodes.

Differentiating $\chi^{2}$ with respect to the zero point $\mu_i$ and setting the derivative to zero gives, for each unit $i$,
\begin{equation}
\sum_{j} n_{ij}\,\left[(\mu_i - \mu_j) - \langle v_i - v_j\rangle\right] = 0,
\label{eq:normal}
\end{equation}
where $n_{ij} \equiv n_{ji}$, $\langle v_j - v_i\rangle \equiv -\langle v_i - v_j\rangle$, and the sum runs over all units $j$ that share at least one cross-unit visit pair with $i$. Collecting terms in $\mu$ yields the linear system
\begin{equation}
A\mu = b,
\label{eq:linsys}
\end{equation}
with
\begin{equation}
\begin{aligned}
A_{ii} &= \sum_{j \neq i} n_{ij}, \\
A_{ij} &= -\,n_{ij} \quad (i \neq j), \\
b_{i} &= \sum_{j \neq i} n_{ij}\,\langle v_i - v_j\rangle .
\end{aligned}
\label{eq:matrix}
\end{equation}
Thus $A_{ii}$ is the total pair count connecting unit $i$ to all other units, $A_{ij}$ encodes the linkage between units $i$ and $j$, and $b_i$ is the corresponding weighted sum of observed velocity differences. Solving $A\mu=b$ yields the zero-point corrections $\mu_i$ for all units simultaneously.

\subsection{Calibration-unit Definitions at Two Levels}
\label{sec:units}

\subsubsection{Exposure level}

The exposure-level partition follows the APOGEE reduction: APRED processes each plate--night with a common wavelength solution and sky/telluric corrections, so we adopt (plate, night) as the minimal exposure-level calibration unit \citep{nidever2015}.

For a chain of units linked only by repeated exposures of a single field, the zero points are constrained only up to an additive constant; in these cases we fix the reference by setting $\mu=0$ for one unit in the chain.

An isolated plate--night unit that shares no cross-unit pairing with any other unit does not enter the matrix, has no solution, and is therefore left uncorrected.

\subsubsection{Fiber level}

The exposure-level correction is a global shift and cannot remove fiber-to-fiber RV offsets, which vary across fibers and evolve with time even within a single night. We therefore define the fiber-level calibration units as (fiber, time window) and solve for these offsets using the exposure-corrected velocities.

We choose the time windows to track instrumental evolution while maintaining sufficient connectivity between units. At APO, the \raisebox{0.5ex}{\texttildelow}2-month summer maintenance shutdown often coincides with comparatively large zero-point changes, so we place window boundaries at observing-year breaks to avoid forcing pre- and post-shutdown data to share a single constant. Windows that are too short, however, leave too few visits (and too few links to other units) to yield reliable solutions; we therefore adopt one observing year per window.

Although LCO has no comparable summer shutdown, we use the same observing-year boundaries there as well: the LCO calibration sample is smaller, and a finer subdivision would leave too few visits per (fiber, window) unit to determine its zero point robustly.

\section{Results}

\subsection{Exposure-level Zero Points}

\begin{figure*}[htbp]
\centering
\includegraphics[width=0.98\textwidth]{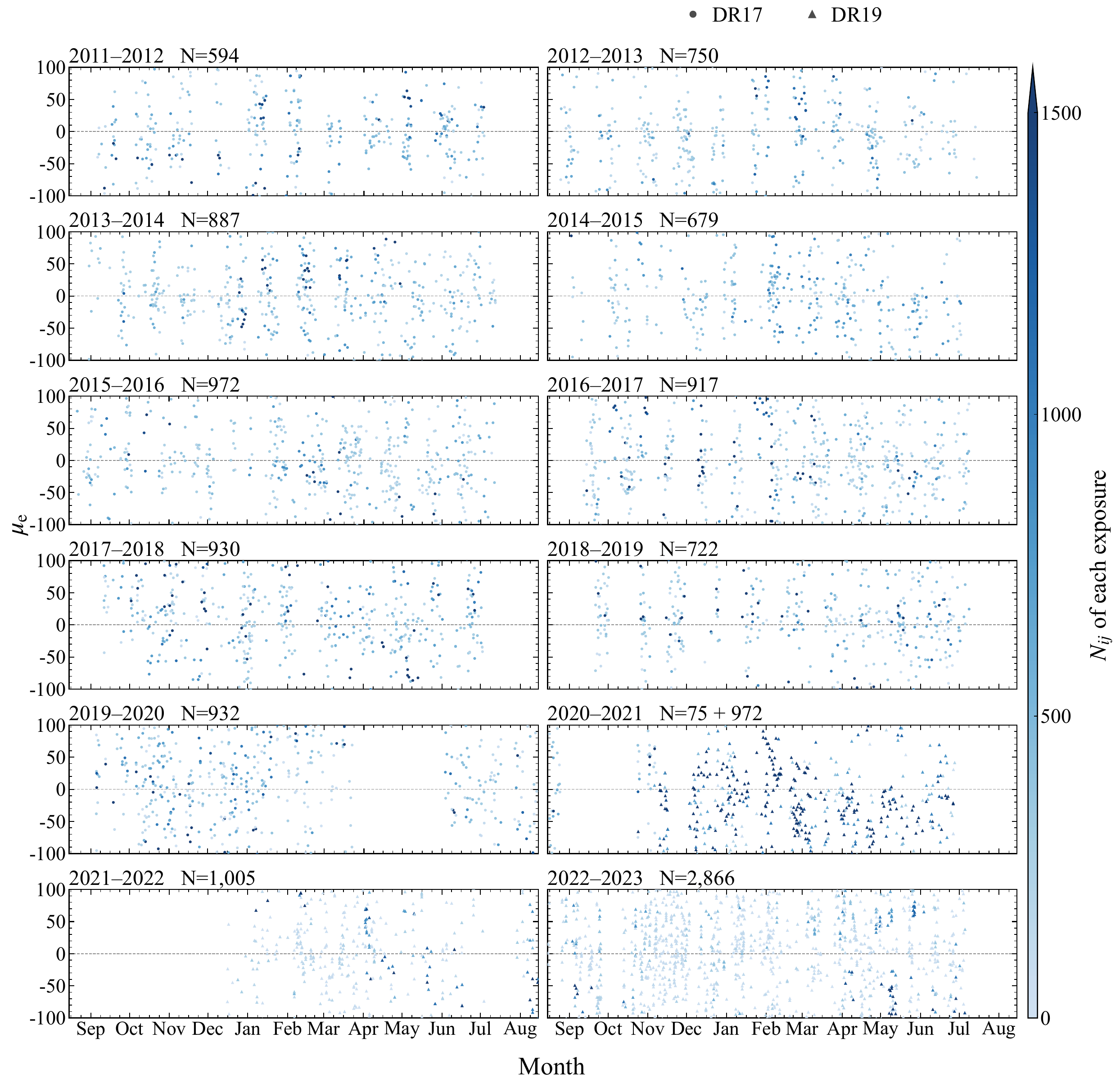}
\caption{Exposure-level zero points $\mu_{\rm e}$ for the APO 2.5~m telescope as a function of time. Each point is one plate--night unit; circles and triangles denote the DR17 and DR19 solutions, respectively. Color encodes $N_{ij}$, the total number of visit pairs summed over all units connected to the unit, indicating the total strength of the repeat-observation constraints.  Each panel spans August~15 of a given year through August~15 of the following year.}
\label{fig:exposure_zp_apo}
\end{figure*}

\begin{figure}[htbp]
\centering
\includegraphics[width=\columnwidth]{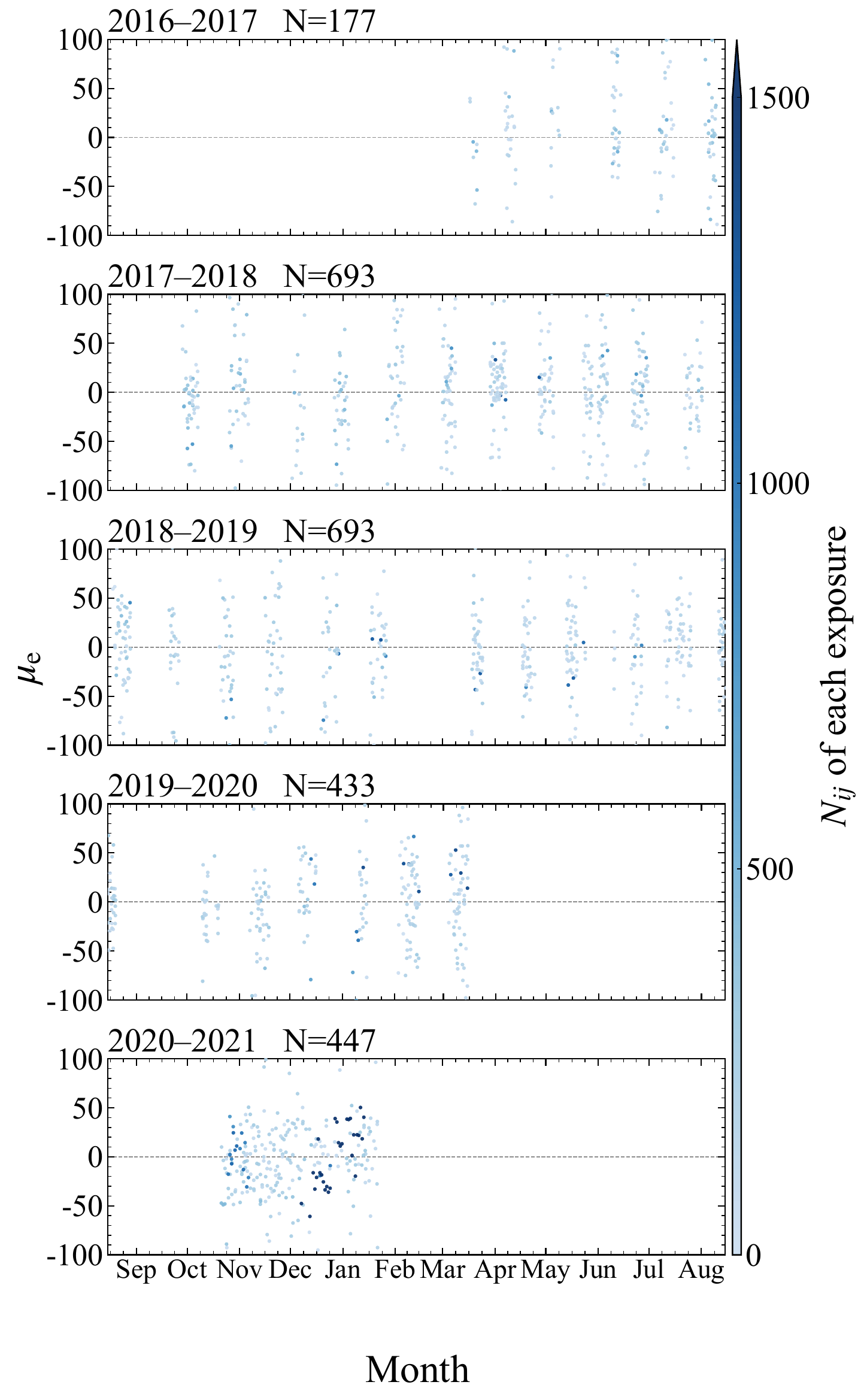}
\caption{Exposure-level zero points $\mu_{\rm e}$ for the LCO 2.5~m telescope (DR17). Symbols and colors are as in Figure~\ref{fig:exposure_zp_apo}. For consistency with the APO figure, panel boundaries are set at August~15.}
\label{fig:exposure_zp_lco}
\end{figure}

We solve for exposure-level zero points for APO DR17 and DR19 jointly, and for LCO DR17 separately. Treating the two APO releases as a single problem is justified for two reasons: (i) 35,039 calibration stars are observed in both releases, providing 424,117 cross-release visit pairs that tightly constrain their relative zero point; and (ii) both releases were obtained with the same spectrograph on the same telescope. LCO, taken with a different spectrograph at a different site, is therefore solved independently. The median absolute corrections among solved exposure units are 91.5 and 157.6~$\mathrm{m\,s^{-1}}$ over the DR17 and DR19 nodes of the joint APO solution, and 46.0~$\mathrm{m\,s^{-1}}$ for LCO DR17, implying typical exposure-level corrections of $\sim$100~$\mathrm{m\,s^{-1}}$ at APO and $\sim$50~$\mathrm{m\,s^{-1}}$ at LCO. Figures~\ref{fig:exposure_zp_apo} and \ref{fig:exposure_zp_lco} show the resulting solutions over time. No zero point could be determined for 2.5\% of exposure units (3.8\% of visits), either because they contain no qualifying calibration visits or because they share no repeat-observed stars with other units and thus provide no cross-unit constraints. We leave these visits uncorrected at the exposure level, exclude them from the fiber-level solution, and flag them in the final catalog with \texttt{RV\_FLAG}$\,=2$.

\subsection{Fiber-level Zero Points}

\begin{figure}[htbp]
\centering
\includegraphics[width=\columnwidth]{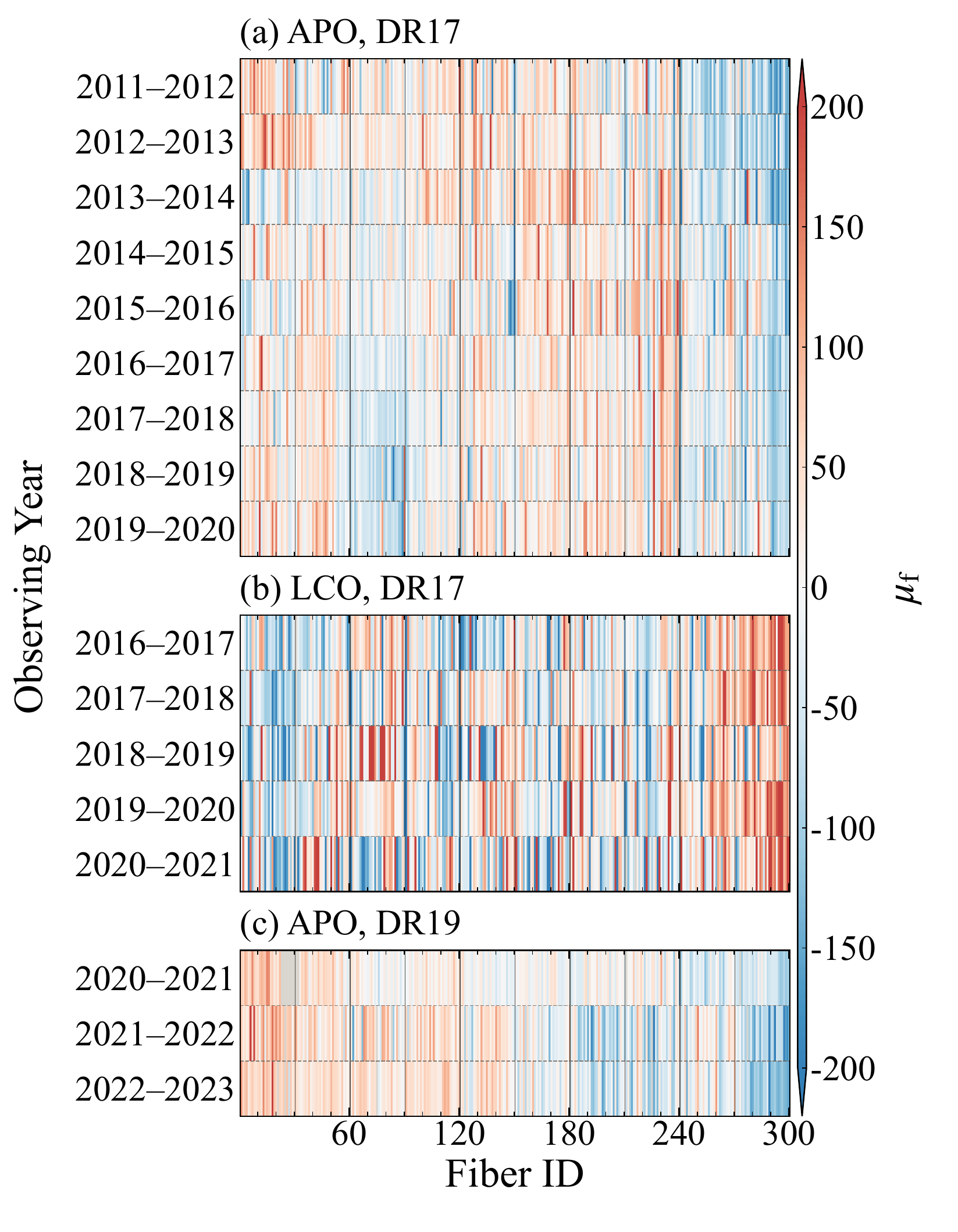}
\caption{\textit{(a)} Fiber--observing-year zero points $\mu_{\rm f}$ for APO 2.5~m DR17 after applying the exposure-level correction. Color shows $\mu_{\rm f}$ ($\mathrm{m\,s^{-1}}$); rows are observing years (dashed horizontal separators) and columns are Fiber ID (solid vertical lines every 30 fibers mark V-groove blocks). The nine rows correspond to the observing years separated by the APO summer gaps. \textit{(b)} Same as \textit{(a)} for LCO 2.5~m DR17. \textit{(c)} Same as \textit{(a)} for APO 2.5~m DR19. All panels share the same color scale ($-200$ to $+200~\mathrm{m\,s^{-1}}$).}
\label{fig:fiber_zp_maps}
\end{figure}

After applying the exposure-level correction, we solve for fiber-level zero points using the fiber--observing-year units defined in Section~3, solving APO DR17 and DR19 jointly and LCO DR17 separately. For APO, time-window boundaries are placed at long gaps in the calibration-night sequence, separating instrumental states across the summer shutdown while preserving adequate sample sizes and cross-unit connectivity; for LCO, we adopt the same observing-year boundaries to keep enough visits per unit. Long-baseline repeat observations then provide cross-constraints between fibers and observing years. The 2,695 DR17 and 878 DR19 fiber--observing-year units are solved together and form one connected network of 3,573 units, with each unit linked to a median of 199 others; LCO DR17, solved on its own, has only 1,486 solved units with a median connectivity of 58. These sparser linkages weaken the uber-calibration constraints and make the LCO fiber-level correction less effective than at APO. The median absolute fiber corrections are 39.7 and 42.3~$\mathrm{m\,s^{-1}}$ for the DR17 and DR19 windows of the joint APO solution, and 65.1~$\mathrm{m\,s^{-1}}$ for LCO DR17. In Figure~\ref{fig:fiber_zp_maps}, persistent vertical stripes indicate a stable fiber-dependent component, while row-to-row changes trace evolution with observing year; bundled structure with Fiber ID reflects the modular slithead, with 30 fibers per V-groove block \citep{wilson2019}.

\clearpage

\section{Verification}

\subsection{Internal repeatability from repeat observations}

\begin{figure*}[htbp]
\centering
\includegraphics[width=0.98\textwidth]{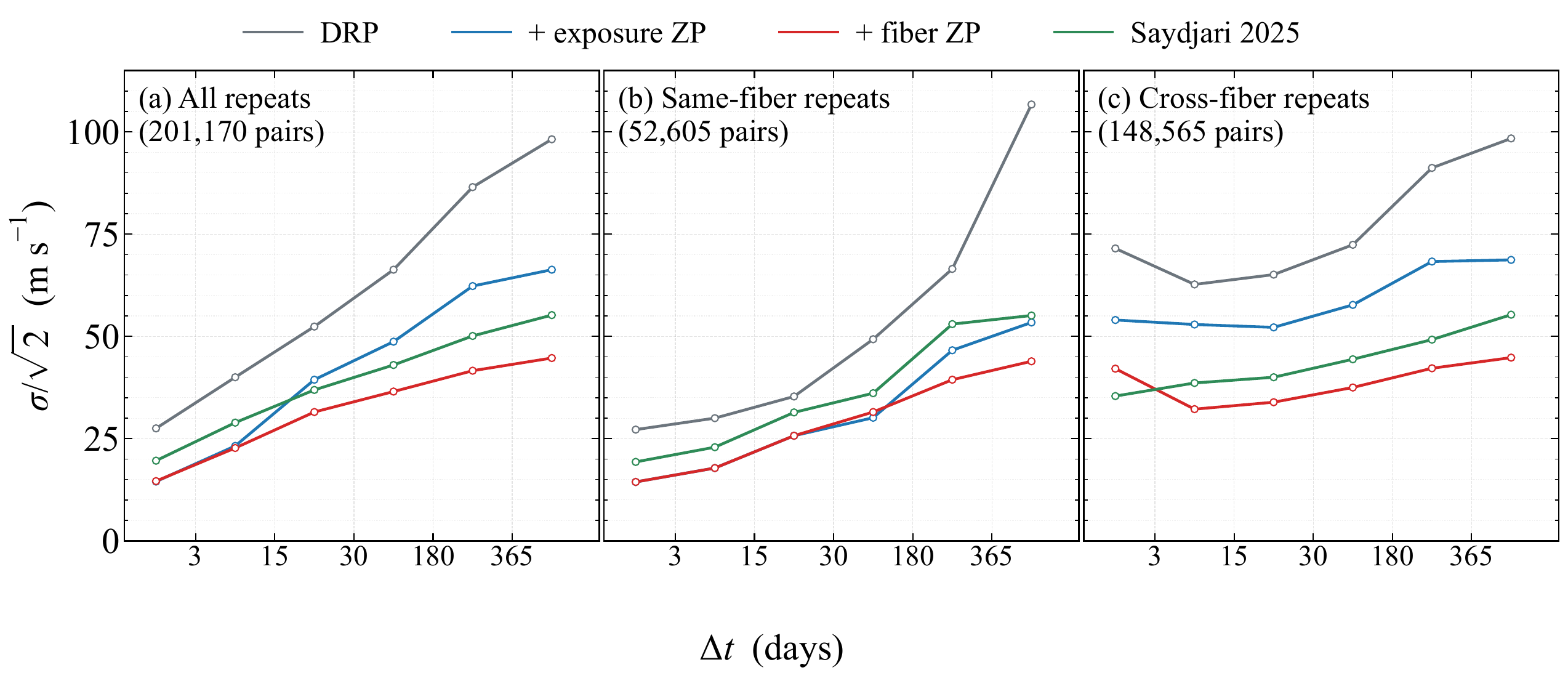}
\caption{Equivalent single-epoch scatter $\sigma_{\rm single}=\sigma(\Delta RV)/\sqrt{2}$ for APO 2.5~m DR17 repeat observations as a function of time separation $\Delta t$. Here $\sigma(\Delta RV)$ is the width of a Gaussian fit to the main peak of the $\Delta RV$ distribution, obtained with $3\sigma$ clipping. Left to right, panels show all repeats, same-fiber repeats, and cross-fiber repeats. Gray, blue, and red curves correspond to the original DRP velocities, the exposure-only correction, and the exposure+fiber correction; the green curve shows \citet{saydjari2025} as an external reference. Points are plotted within each $\Delta t$ interval; tick labels at 3, 15, 30, 180, and 365~days indicate interval boundaries (leftmost: $1\leq\Delta t<3$~days; rightmost: $\Delta t\geq365$~days). We require ${\rm S/N}_{\min}>150$ for all pairs. Because same-fiber visits within the same observing year share the same fiber zero point, which cancels in $\Delta RV$, parts of the blue and red curves coincide in the same-fiber panel.}

\label{fig:repeat_dt_apo}
\end{figure*}

We quantify internal repeatability using the visit-to-visit radial-velocity difference $\Delta RV$ for the same star, and define the equivalent single-epoch scatter $\sigma_{\rm single}$ as the width of a Gaussian fit to the main peak of the $\Delta RV$ distribution divided by $\sqrt{2}$. We bin pairs by time separation $\Delta t$, choosing bin boundaries where the fraction of same-fiber pairs changes in discrete steps. This yields 6 intervals: $\le 2$~days, 3--15~days, 15--30~days, 30--180~days, 180--365~days, and $>365$~days. Figure~\ref{fig:repeat_dt_apo} reports all repeats, same-fiber repeats, and cross-fiber repeats in each interval, and compares the original DRP velocities with the exposure-only and exposure+fiber corrections; the results of \citet{saydjari2025} are shown for external reference.

\begin{figure*}[htbp]
\centering
\includegraphics[height=0.87\textheight]{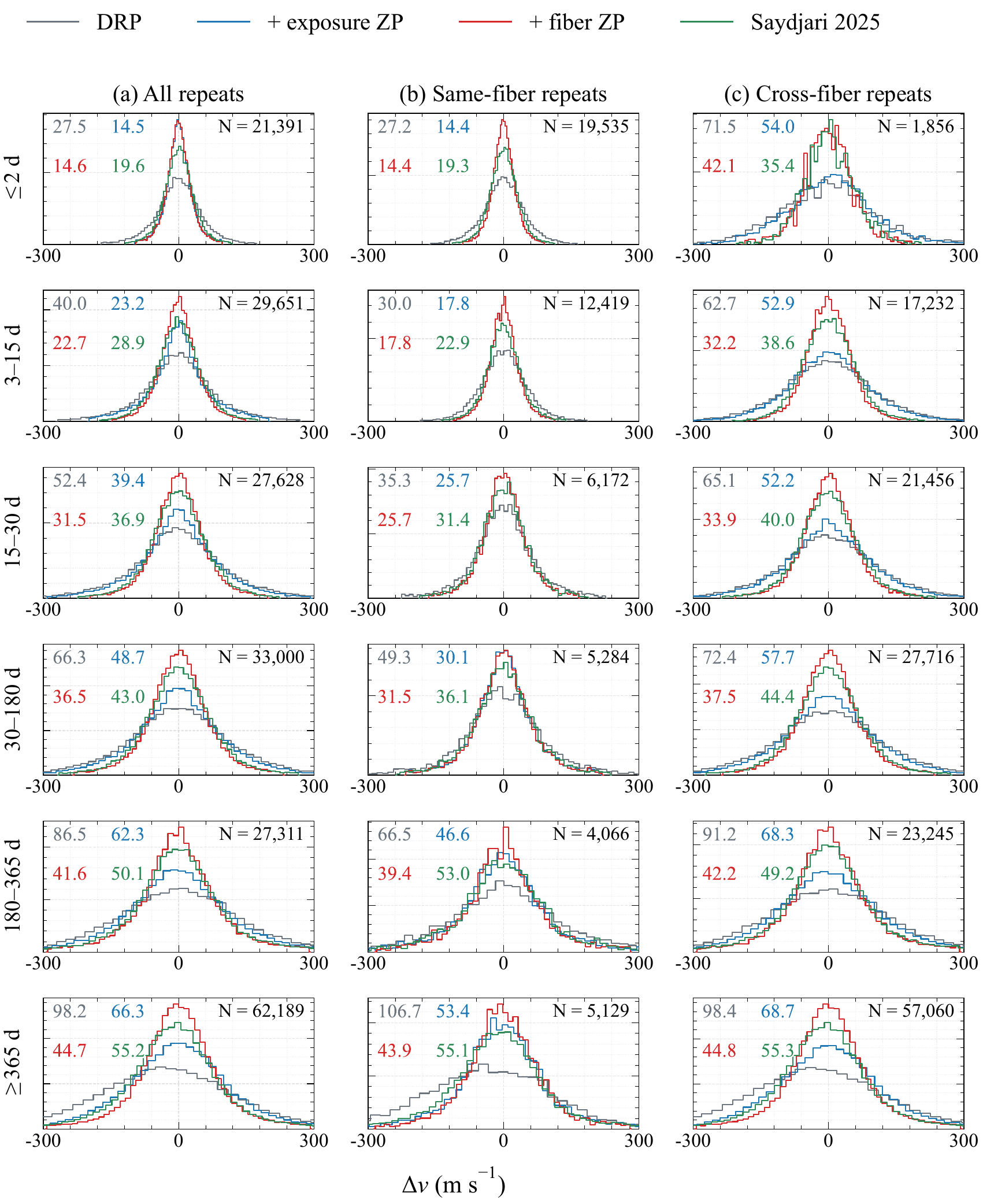}
\caption{$\Delta RV$ distributions for APO 2.5~m DR17 repeat observations. Rows (top to bottom) show the six $\Delta t$ intervals: $\le 2$~days, 3--15~days, 15--30~days, 30--180~days, 180--365~days, and $>365$~days; columns (left to right) show all repeats, same-fiber repeats, and cross-fiber repeats. Curve colors match Figure~\ref{fig:repeat_dt_apo}. In each cell, the colored number at upper left gives $\sigma_{\rm single}$ (in $\mathrm{m\,s^{-1}}$) for that curve, and $N$ at upper right gives the number of repeat pairs.}

\label{fig:repeat_hist_apo}
\end{figure*}

For repeats with $\Delta t>365$~days, $\sigma_{\rm single}$ drops from 98.2~$\mathrm{m\,s^{-1}}$ (original DRP velocities) to 66.3~$\mathrm{m\,s^{-1}}$ after the exposure-level correction, and to 44.7~$\mathrm{m\,s^{-1}}$ after the additional fiber-level correction. The exposure-level correction reduces the scatter in all $\Delta t$ bins, with the largest gains at long baselines. Splitting by same-fiber versus cross-fiber repeats shows that the incremental improvement from the fiber-level correction is driven primarily by cross-fiber pairs: for cross-fiber repeats with $\Delta t>365$~days, $\sigma_{\rm single}$ decreases from 68.7 to 44.8~$\mathrm{m\,s^{-1}}$. Figure~\ref{fig:repeat_hist_apo} shows the corresponding $\Delta RV$ distributions; their main peaks narrow after correction, most strongly for long-baseline and cross-fiber pairs, demonstrating improved internal consistency.

We apply the same $\Delta t$ binning and fiber-pairing classification to APO DR19 and LCO DR17, and compare the repeat-observation scatter before and after correction within each data set.

\begin{figure*}[htbp]
\centering
\includegraphics[width=0.98\textwidth]{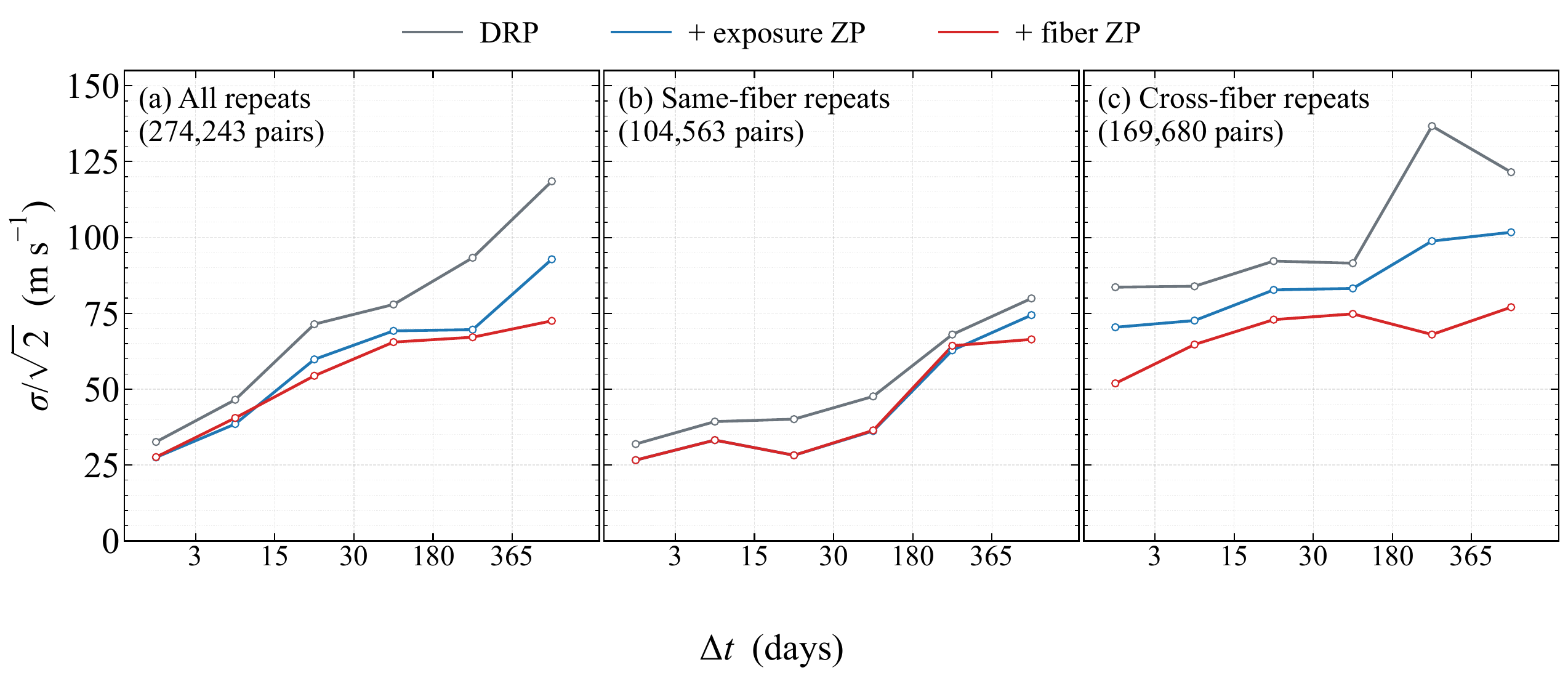}
\caption{Equivalent single-epoch scatter $\sigma_{\rm single}$ for APO 2.5~m DR19 repeat observations as a function of time separation $\Delta t$. Panels, $\Delta t$ binning, axis reading, and S/N thresholds match Figure~\ref{fig:repeat_dt_apo}. No external reference curve is shown because \citet{saydjari2025} does not cover DR19.}
\label{fig:repeat_dt_apo_dr19}
\end{figure*}

\begin{figure*}[htbp]
\centering
\includegraphics[height=0.87\textheight]{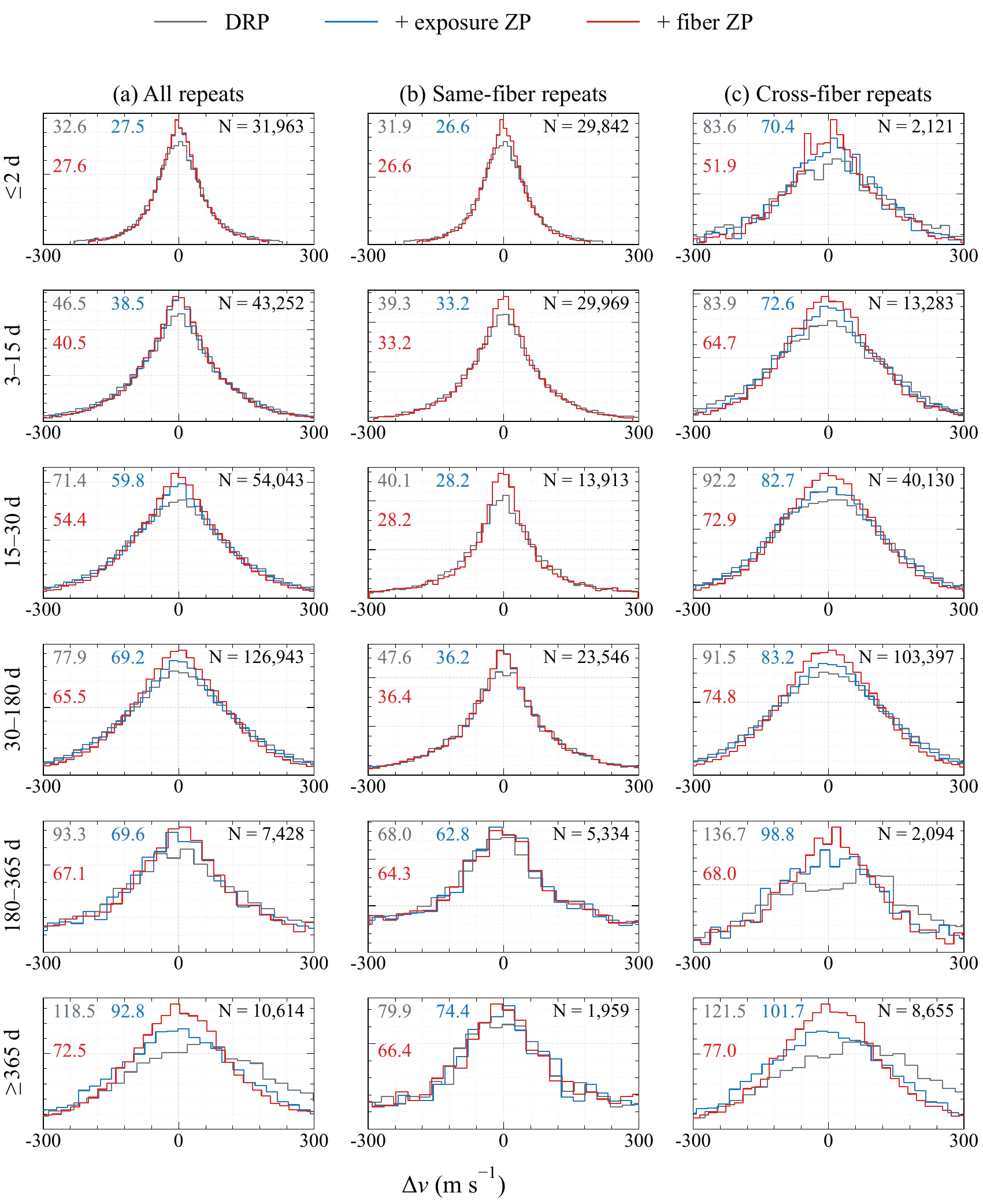}
\caption{$\Delta RV$ distributions for APO 2.5~m DR19 repeat observations. The row/column layout and the meanings of the colored $\sigma_{\rm single}$ and pair count $N$ match Figure~\ref{fig:repeat_hist_apo}; curve colors match Figure~\ref{fig:repeat_dt_apo_dr19}. In the same-fiber column, the blue and red curves coincide at short baselines because both visits fall within the same observing-year window and the fiber zero point cancels in $\Delta RV$; only cross-window pairs retain sensitivity to the fiber-level correction. No comparison curve from \citet{saydjari2025} is available for DR19.}
\label{fig:repeat_hist_apo_dr19}
\end{figure*}

For APO DR19 repeats with $\Delta t>365$~days, $\sigma_{\rm single}$ decreases from 118.5~$\mathrm{m\,s^{-1}}$ (original DRP velocities) to 92.8~$\mathrm{m\,s^{-1}}$ after the exposure-level correction, and to 72.5~$\mathrm{m\,s^{-1}}$ after the additional fiber-level correction. The exposure-level correction reduces the scatter in every $\Delta t$ bin across all three pairing classes. The incremental gain from the fiber-level correction is concentrated at long baselines: for $\Delta t>365$~days, $\sigma_{\rm single}$ decreases from 74.4 to 66.4~$\mathrm{m\,s^{-1}}$ for same-fiber repeats and from 101.7 to 77.0~$\mathrm{m\,s^{-1}}$ for cross-fiber repeats, while some shorter-baseline bins show little change or a slight increase. We therefore interpret the DR19 fiber-level correction as being most robust on long, cross-observing-year time scales. Figure~\ref{fig:repeat_hist_apo_dr19} supports this picture: the long-baseline $\Delta RV$ peaks narrow strongly after the exposure-level correction and tighten further once the fiber zero points are included.

DR19 straddles two observing eras, the plug-plate era and the FPS era. In the plug-plate era the fiber through which a star is observed is essentially randomly assigned from one visit to the next, so the release accumulates a large number of cross-fiber repeat observations; after the transition to the FPS, robotic fiber positioning places about 90\% of the repeat observations of a given star on the same fiber. The constraints on the fiber-level correction, and the gain from it, therefore come mainly from cross-fiber repeats within the plug-plate era and from repeats that straddle the plug-plate and FPS eras.

Precisely because DR19 contains both observing modes, repeats with long time separations frequently straddle the two eras, which makes the long-timescale consistency of the original DRP velocities markedly worse in DR19 than in DR17. Once APO DR17 and DR19 are solved jointly, the two releases are placed on a common velocity zero-point gauge. For stars observed in both releases, the equivalent single-epoch scatter of the cross-release repeat velocity differences (pairs with $\Delta t>365$~days and ${\rm S/N}_{\min}>150$) decreases from 96.0~$\mathrm{m\,s^{-1}}$ for the original DRP velocities to 63.9~$\mathrm{m\,s^{-1}}$, a substantial improvement in the velocity consistency between the two releases.

\begin{figure*}[htbp]
\centering
\includegraphics[width=0.98\textwidth]{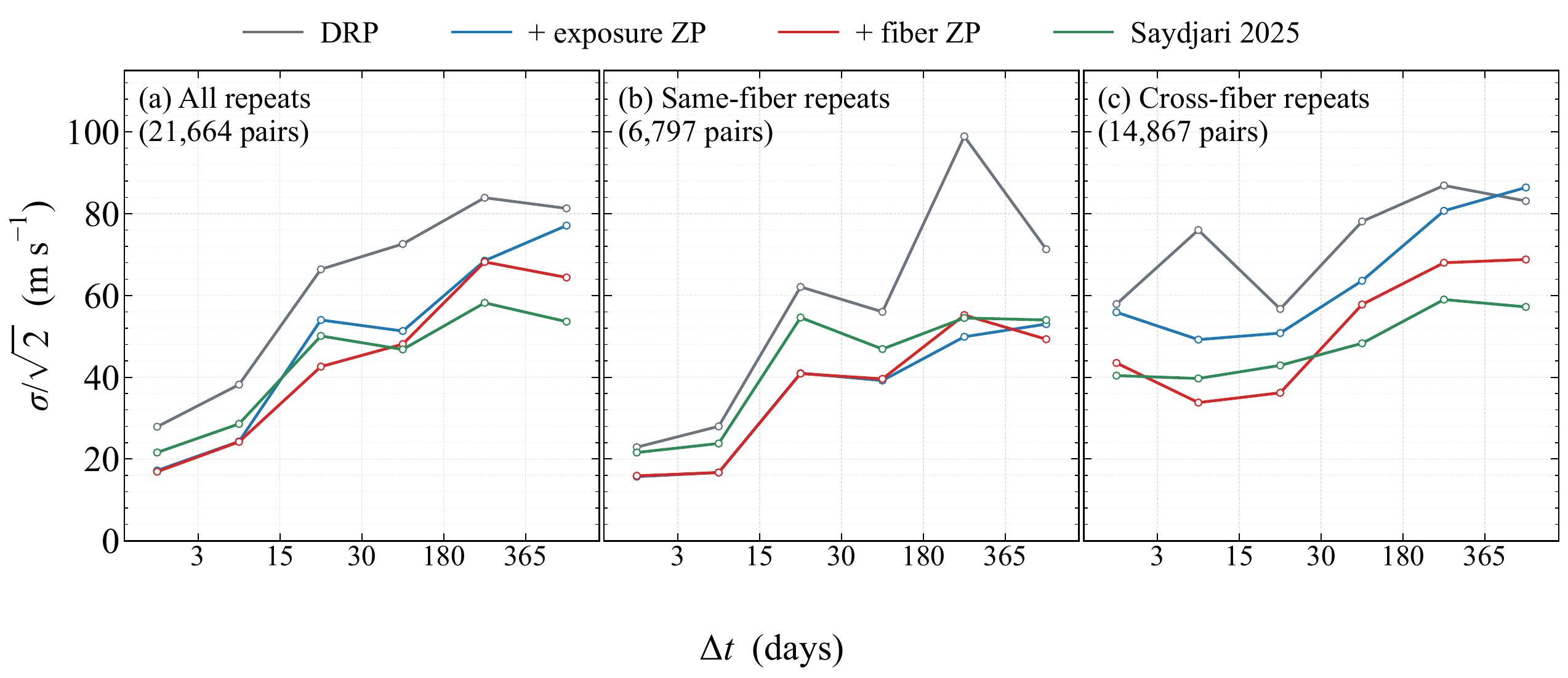}
\caption{Equivalent single-epoch scatter $\sigma_{\rm single}$ for LCO 2.5~m DR17 repeat observations as a function of time separation $\Delta t$. Panels, $\Delta t$ binning, axis reading, and S/N thresholds match Figure~\ref{fig:repeat_dt_apo}. Gray, blue, and red curves show the original DRP velocities, the exposure-only correction, and the exposure+fiber correction (using LCO observing-year fiber zero points); the green curve shows \citet{saydjari2025} as an external reference.}

\label{fig:repeat_dt_lco}
\end{figure*}

\begin{figure*}[htbp]
\centering
\includegraphics[height=0.87\textheight]{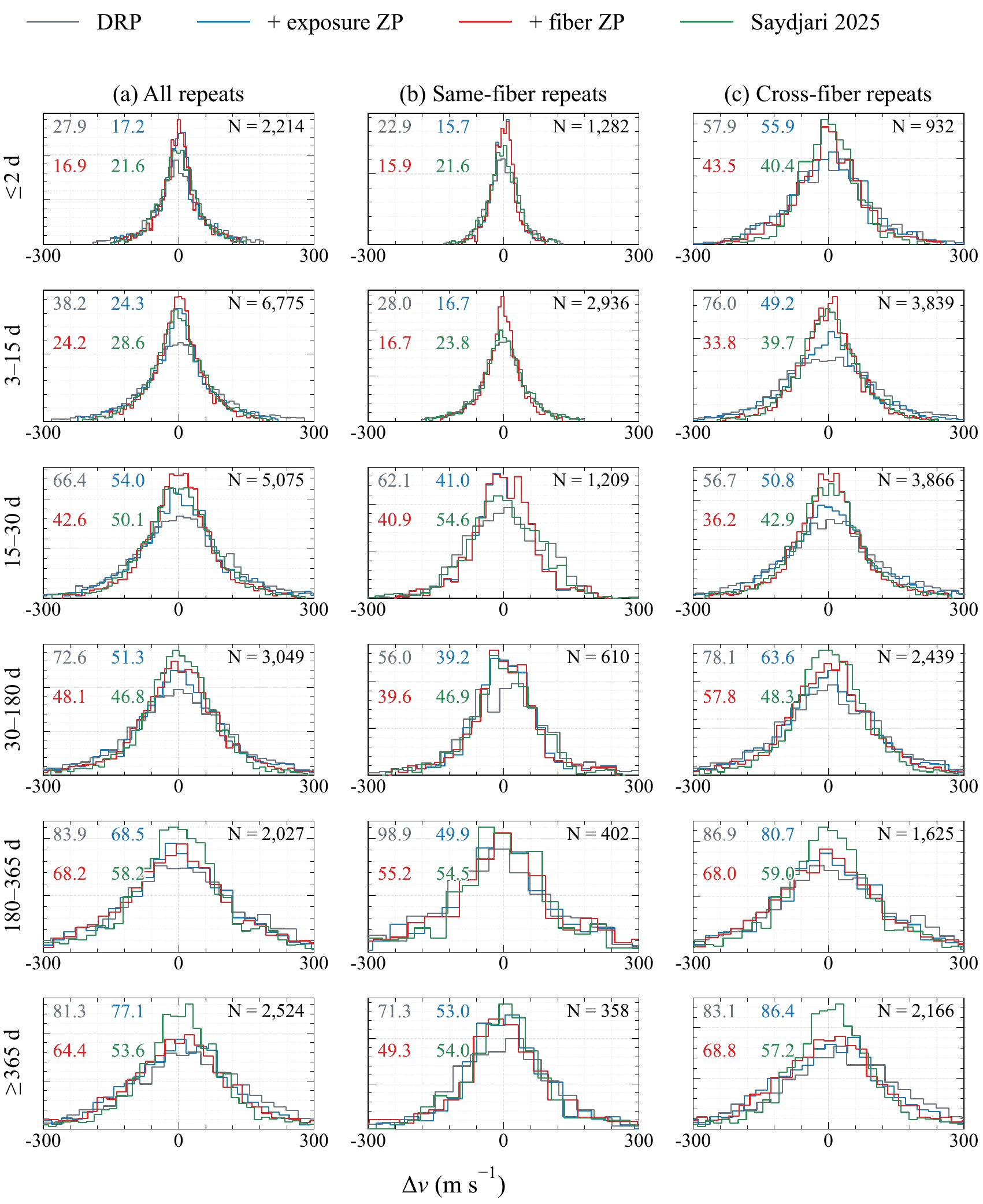}
\caption{$\Delta RV$ distributions for LCO 2.5~m DR17 repeat observations. The row/column layout and the meanings of the colored $\sigma_{\rm single}$ and pair count $N$ match Figure~\ref{fig:repeat_hist_apo}; curve colors match Figure~\ref{fig:repeat_dt_lco}. The colored $\sigma_{\rm single}$ values are adopted from Figure~\ref{fig:repeat_dt_lco} (not refit from the histograms shown here), so this figure is intended for comparing distribution shapes and widths rather than alternative scatter estimators.}

\label{fig:repeat_hist_lco}
\end{figure*}

For LCO DR17 repeats with $\Delta t>365$~days, $\sigma_{\rm single}$ decreases from 81.3~$\mathrm{m\,s^{-1}}$ (original DRP velocities) to 77.1~$\mathrm{m\,s^{-1}}$ after the exposure-level correction, and to 64.4~$\mathrm{m\,s^{-1}}$ after the additional fiber-level correction; in the 15--30~day bin it falls from 66.4 to 54.0 to 42.6~$\mathrm{m\,s^{-1}}$. Splitting by same-fiber versus cross-fiber repeats again shows that the fiber-level gain is driven primarily by cross-fiber pairs: for cross-fiber repeats with $\Delta t>365$~days, $\sigma_{\rm single}$ drops from 86.4 to 68.8~$\mathrm{m\,s^{-1}}$. Figure~\ref{fig:repeat_hist_lco} shows the corresponding $\Delta RV$ distributions; where the improvement is clearest, the main peak narrows after the two-level correction, indicating improved internal consistency of the LCO repeat observations.

The LCO corrections are smaller than for APO DR17 because the calibration network is much sparser. The LCO calibration sample is only $\sim$24\% of APO's, with fewer calibration visits per unit (median 65 vs.~107) and fewer repeat-observation links (median 6 edges per unit vs.~12), which increases the zero-point solution noise and reduces the impact of the exposure-level correction (81.3$\to$77.1~$\mathrm{m\,s^{-1}}$ at LCO vs.~98.2$\to$66.3~$\mathrm{m\,s^{-1}}$ at APO for $\Delta t>365$~days).

The fiber-level network is sparser still: LCO has 1,493 fiber--observing-year units with median connectivity 57 and an average of 288 cross-unit repeat pairs (vs.~2,695 units, 184, and 944 for APO). Only 34.9\% of LCO fiber pairs are directly linked by repeats (median 2 shared repeat pairs per fiber pair), compared with 77.1\% (median 7) for APO, limiting the precision of the LCO fiber zero points and thus the achievable improvement.

\subsection{APO--LCO Cross-site Consistency}

\begin{figure}[htbp]
\centering
\includegraphics[width=\columnwidth]{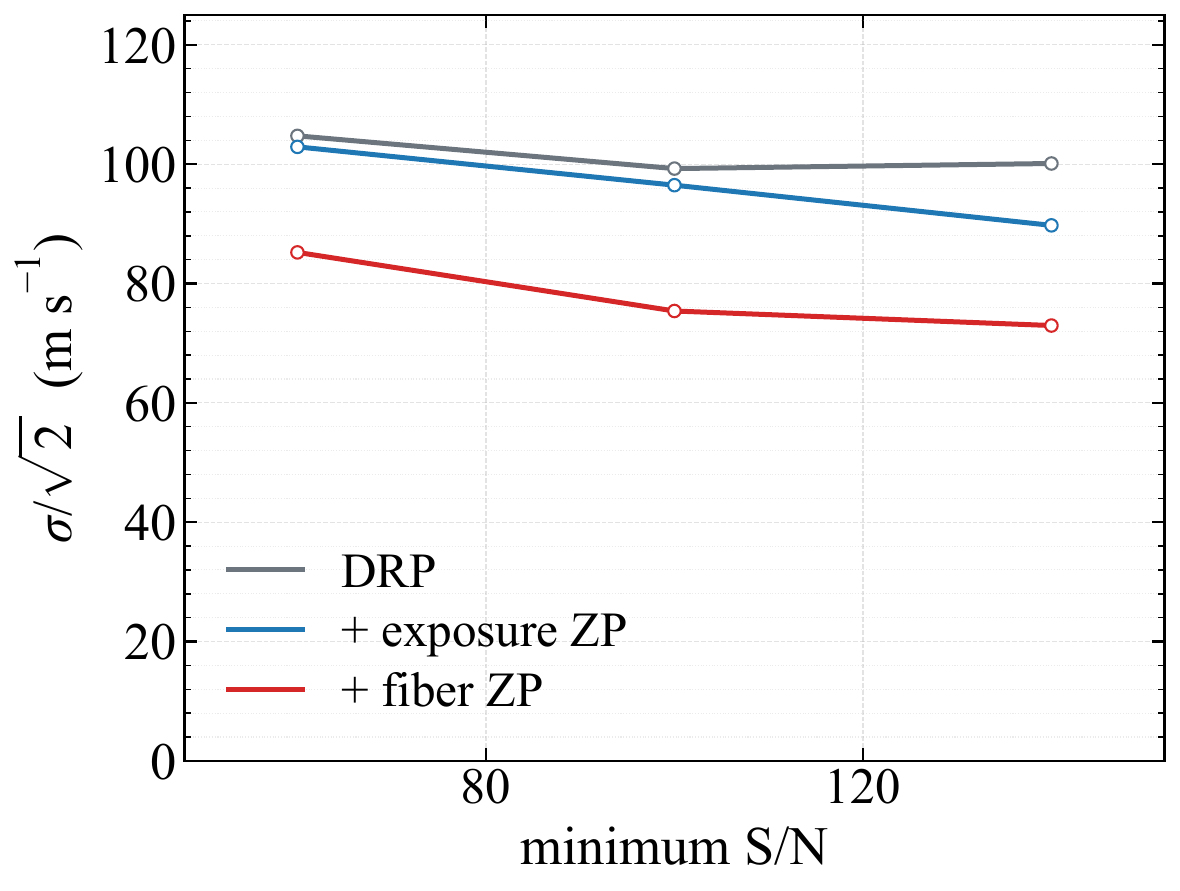}
\caption{Cross-site scatter for DR17 sources observed in common by APO and LCO with $\Delta t>365$~days. For each star we define $\Delta v\equiv v_{\rm APO}-v_{\rm LCO}$ and, within each S/N bin, subtract the median $\Delta v$ before computing the scatter (the subtracted constant is typically $-130$ to $-150~\mathrm{m\,s^{-1}}$). The vertical axis shows the equivalent single-visit scatter, $\sigma(\Delta v)/\sqrt{2}$. The three points correspond to ${\rm S/N}_{\min}=40$--80, 80--120, and $\ge120$; 80 and 120 mark bin boundaries, and the highest-S/N bin is open-ended. Gray, blue, and red curves show the original DRP velocities, the exposure-only correction, and the exposure+fiber correction, respectively.}
\label{fig:apo_lco_common}
\end{figure}

APO and LCO use different spectrographs, so their exposure- and fiber-level zero points are solved independently. Using DR17 stars observed by both telescopes, we form all APO--LCO visit pairs for each star and restrict to pairs with $\Delta t>365$~days to probe cross-observing-year consistency of the fiber-level correction. This yields 17,076 pairs from 831 stars. Binning by ${\rm S/N}_{\min}$ (40--80, 80--120, and $\ge120$), the equivalent single-visit scatter decreases from 104.7$\to$102.9$\to$85.2, 99.3$\to$96.5$\to$75.4, and 100.1$\to$89.7$\to$73.0~$\mathrm{m\,s^{-1}}$, respectively (original, exposure-only, exposure+fiber; Figure~\ref{fig:apo_lco_common}). Thus, after the full correction the APO--LCO differences are less scattered than in the original DRP in all S/N bins, with the largest improvement at high S/N, indicating improved cross-site consistency for the common-source sample.

\subsection{Uncertainty Estimation}
\label{sec:uncertainty}
\begin{figure*}[htbp]
\centering
\includegraphics[width=0.98\textwidth]{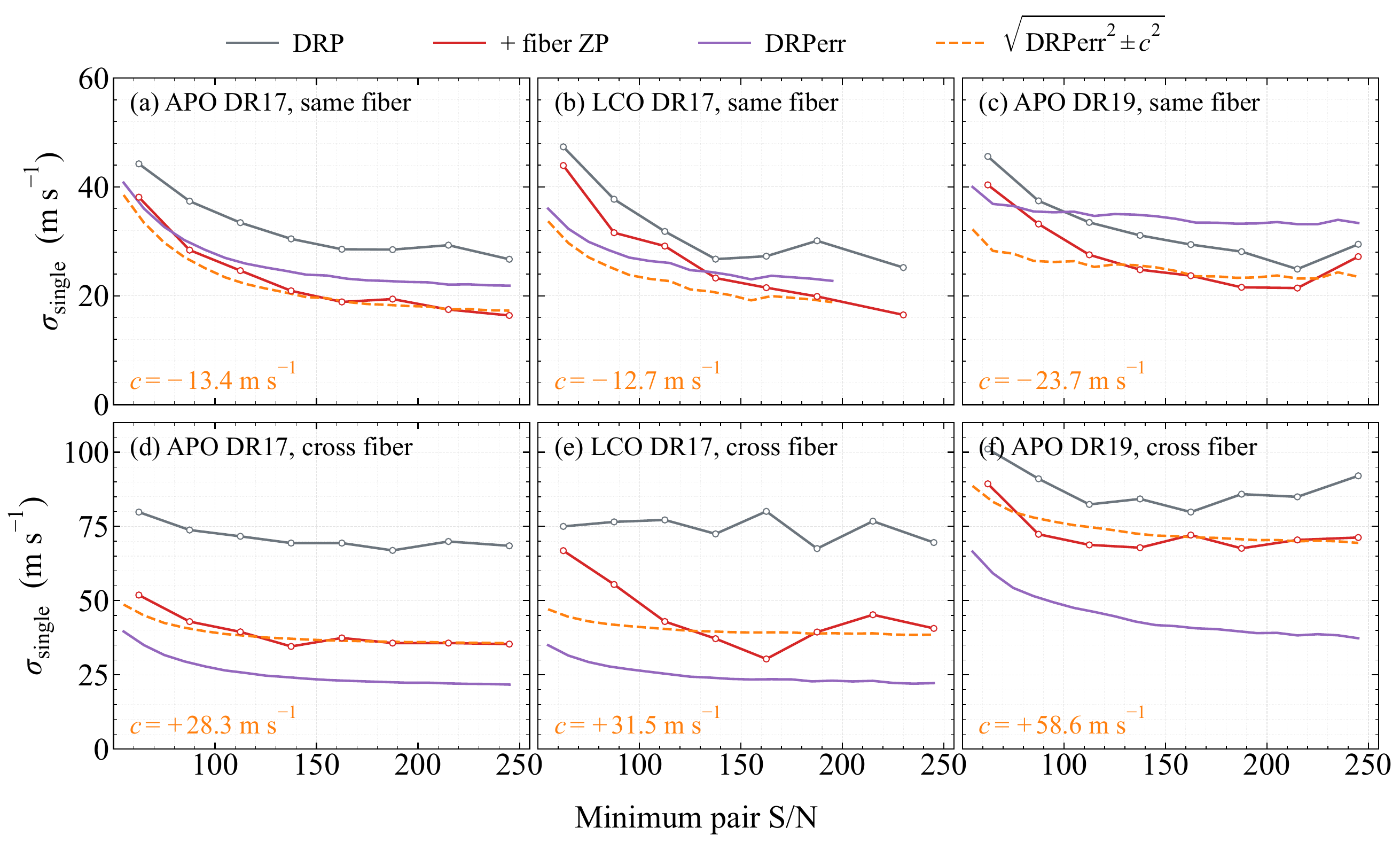}
\caption{Empirical calibration of the reported velocity uncertainty at high signal-to-noise. Panels show APO DR17, LCO DR17, and APO DR19, for repeat pairs taken through the same fiber \textit{(a)--(c)} and through different fibers \textit{(d)--(f)}; note the different vertical scales of the two rows. In each panel, the gray curve is the equivalent single-visit scatter $\sigma_{\rm single}$ inferred from repeat-visit velocities in the original DRP output, the red curve shows $\sigma_{\rm single}$ after the two-level zero-point correction, and the purple curve is the median DRP formal error \texttt{VRELERR} of the individual visits in those repeat pairs, binned by each visit's ${\rm S/N}$. Repeat pairs are restricted to $\Delta t<365$~days and to visit pairs whose ${\rm S/N}$ agree to within 25\% of the lower of the two; DR19 plug-plate and FPS data are combined. The orange dashed curve shows the calibrated per-visit uncertainty, $\sqrt{\texttt{VRELERR}^{2}\pm c^{2}}$, where the term $c$ (printed in each panel) is fitted by equal-weight least squares to the red curve over the ${\rm S/N}$ bins between 150 and 250 (points plotted at bin centers; the highest bin is open-ended and drawn at a nominal abscissa). In panel \textit{(b)} the last two bins are merged, as the same-fiber LCO sample is too sparse above ${\rm S/N}=200$.}
\label{fig:error_calibration}
\end{figure*}
To assess whether the catalog velocity uncertainties match the empirical repeatability of the measurements, we compare three high-${\rm S/N}$ curves: the DRP formal velocity error \texttt{VRELERR}; the repeat-observation scatter of the original velocities \texttt{VHELIO}; and the repeat-observation scatter of the corrected velocities \texttt{VHELIO\_CAL} derived in this work.These three curves are shown in Figure~\ref{fig:error_calibration} for each of the three data sets, separately for same-fiber and cross-fiber repeats.

For the repeat-observation sample, we compute the original and corrected velocity differences separately and define the equivalent single-visit scatter as the width of a Gaussian fit to the main peak of the velocity-difference distribution in each ${\rm S/N}$ bin, divided by $\sqrt{2}$. For the DRP formal uncertainty, we take all individual visits contributing to these repeat-observation statistics, bin them by each visit's ${\rm S/N}$, and adopt the median \texttt{VRELERR} per bin. To ensure the repeat-observation scatter primarily traces time-variable errors within a single observing system, we require the two visits to be separated by less than one year. We also require their ${\rm S/N}$ to agree to within 25\% of the lower of the two, so that a pair is not binned at a high ${\rm S/N}$ while its scatter is set by a much fainter visit. We further split the repeat pairs according to whether the two visits were taken through the same fiber, since the fiber-level zero point cancels in same-fiber pairs but not in cross-fiber ones.

Comparing the repeat-observation scatter of the original \texttt{VHELIO} to \texttt{VRELERR}, we find that for cross-fiber repeats in all three data sets the DRP formal errors lie well below the empirical scatter of the original velocities. This suggests that \texttt{VRELERR} largely reflects the statistical measurement error under ideal conditions and does not fully capture systematic contributions (e.g., zero-point drift), thereby underestimating the true uncertainty. In contrast, for same-fiber repeats the original repeat-observation scatter is close to \texttt{VRELERR}, implying comparatively weak systematics once the fiber-to-fiber term is removed by construction.

After applying the zero-point corrections presented in this work, the repeat-observation scatter decreases markedly in the three cross-fiber samples dominated by systematics, indicating that the corrections remove most of the systematic component in the original velocities. The corrected scatter is still not fully consistent with \texttt{VRELERR}: it remains slightly higher for cross-fiber repeats in APO DR17 and LCO DR17, and is still appreciably higher for cross-fiber repeats in DR19. This implies residual errors that are neither fully captured by the current correction model nor represented in the DRP formal uncertainties. Conversely, for same-fiber repeats the corrected scatter is slightly below \texttt{VRELERR}, suggesting that the formal errors are somewhat conservative in that regime.

We adopt a simple empirical calibration: at high ${\rm S/N}$ we introduce an ${\rm S/N}$-independent correction term $c$ and combine it with the DRP formal uncertainty in variance space. For same-fiber repeats we obtain
\begin{equation}
\label{eq:cerr_same}
\begin{aligned}
c_{\mathrm{DR17,APO,same}} &= -13.4~\mathrm{m\,s^{-1}}, \\
c_{\mathrm{DR17,LCO,same}} &= -12.7~\mathrm{m\,s^{-1}}, \\
c_{\mathrm{DR19,APO,same}} &= -23.7~\mathrm{m\,s^{-1}}.
\end{aligned}
\end{equation}
For cross-fiber repeats, which retain the residual fiber-to-fiber term, we find
\begin{equation}
\label{eq:cerr_cross}
\begin{aligned}
c_{\mathrm{DR17,APO,cross}} &= +28.3~\mathrm{m\,s^{-1}}, \\
c_{\mathrm{DR17,LCO,cross}} &= +31.5~\mathrm{m\,s^{-1}}, \\
c_{\mathrm{DR19,APO,cross}} &= +58.6~\mathrm{m\,s^{-1}}.
\end{aligned}
\end{equation}
For an observing period $i$ and each of the two fiber cases, we define the corrected single-visit velocity uncertainty as
\begin{equation}
\label{eq:sigma_corr}
\sigma_{\mathrm{RV,corr},i}=
\begin{cases}
\sqrt{\mathrm{VRELERR}^{2}+c_i^{2}}, & c_i\geq 0,\\[4pt]
\sqrt{\mathrm{VRELERR}^{2}-c_i^{2}}, & c_i<0,
\end{cases}
\end{equation}
where $c_i>0$ indicates that \texttt{VRELERR} underestimates the observed scatter (requiring an additional term in quadrature), while $c_i<0$ indicates that \texttt{VRELERR} is conservative (and can be reduced in variance space). We release both cases as separate columns, so that users can adopt the one matching their science goal. For the same-fiber cases, $c<0$ implies that the expression is undefined when \texttt{VRELERR}$<|c|$, and yields unrealistically small values near that threshold; we therefore impose a floor of $15~\mathrm{m\,s^{-1}}$ on the same-fiber $\sigma_{\mathrm{RV,corr}}$, which affects 6.9\% of the catalog.

\section{Data Product}
We release a per-visit, value-added radial-velocity catalog that combines APOGEE DR17 (APO 2.5~m and LCO 2.5~m) with DR19 (APO 2.5~m), totaling 3,501,727 visits. Of these, 3,336,141 visits receive the full two-level zero-point correction, corresponding to 930,519 unique stars after removing duplicates. The primary corrected-velocity column is \texttt{VHELIO\_CAL}.

To preserve one-to-one correspondence with the source catalogs, we retain all visits, including those that cannot be reliably corrected: 3,336,141 rows have both exposure- and fiber-level corrections, 17,034 have exposure-only corrections, and 148,552 have no correction. These cases are flagged by \texttt{RV\_FLAG}$\,=0$, 1, and 2, respectively; velocity columns for uncorrected rows are set to null. \texttt{VHELIO\_CAL\_ERR\_SAME} and \texttt{VHELIO\_CAL\_ERR\_CROSS} are null for those same rows. Users requiring the full two-level product should select \texttt{RV\_FLAG}$\,=0$. The catalog is publicly available on Zenodo (\dataset[doi:10.5281/zenodo.22007314]{\doi{10.5281/zenodo.22007314}}).

The catalog also includes the target identifier; a pointer back to the source \texttt{allVisit} row; observation time; plate and fiber identifiers; sky position; signal-to-noise ratio; original quality flags; and available stellar parameters and targeting information. All 35 fields are summarized in Table~\ref{tab:catalog_columns}. This product supports long-baseline repeatability studies, selection of radial-velocity variables, and Galactic kinematics.

\newcommand{\coldesc}[1]{\parbox[t]{0.45\textwidth}{\raggedright #1\strut}}
\begin{deluxetable*}{clll}
\tabletypesize{\footnotesize}
\tablecaption{Column description of the value-added per-visit radial-velocity
catalog.\label{tab:catalog_columns}}
\tablewidth{0pt}
\tablehead{
\colhead{Column} & \colhead{Name} & \colhead{Unit} &
\colhead{Description}
}
\startdata
1 & \texttt{RELEASE} & \nodata & \coldesc{Source data release, either \texttt{DR17} or \texttt{DR19}} \\
2 & \texttt{ROW} & \nodata & \coldesc{Zero-based row index in the \texttt{allVisit} file named by \texttt{RELEASE}} \\
3 & \texttt{APOGEE\_ID} & \nodata & \coldesc{Target identifier} \\
4 & \texttt{TELESCOPE} & \nodata & \coldesc{Telescope, \texttt{apo25m} or \texttt{lco25m}} \\
5 & \texttt{FIELD} & \nodata & \coldesc{Name of the observed field} \\
6 & \texttt{PLATE} & \nodata & \coldesc{Plate number; the FPS field identifier for DR19} \\
7 & \texttt{MJD} & d & \coldesc{Modified Julian date of the observing night} \\
8 & \texttt{MJD\_MID} & d & \coldesc{Modified Julian date of the exposure midpoint, converted from the header JD} \\
9 & \texttt{FIBERID} & \nodata & \coldesc{Fiber number} \\
10 & \texttt{RA} & deg & \coldesc{Right ascension (J2000)} \\
11 & \texttt{DEC} & deg & \coldesc{Declination (J2000)} \\
12 & \texttt{GLON} & deg & \coldesc{Galactic longitude} \\
13 & \texttt{GLAT} & deg & \coldesc{Galactic latitude} \\
14 & \texttt{SNR} & \nodata & \coldesc{Combined signal-to-noise ratio of the visit} \\
15 & \texttt{snr\_blue} & \nodata & \coldesc{Signal-to-noise ratio of the blue detector} \\
16 & \texttt{snr\_green} & \nodata & \coldesc{Signal-to-noise ratio of the green detector} \\
17 & \texttt{snr\_red} & \nodata & \coldesc{Signal-to-noise ratio of the red detector} \\
18 & \texttt{VHELIO} & km\,s$^{-1}$ & \coldesc{Heliocentric radial velocity from the original pipeline} \\
19 & \texttt{VREL} & km\,s$^{-1}$ & \coldesc{Relative radial velocity from the original pipeline} \\
20 & \texttt{VRELERR} & km\,s$^{-1}$ & \coldesc{Relative radial-velocity error reported by the original pipeline} \\
21 & \texttt{STARFLAG} & \nodata & \coldesc{Visit-level quality bitmask from the original pipeline} \\
22 & \texttt{VHELIO\_CAL} & km\,s$^{-1}$ & \coldesc{Recommended velocity of this work: the visit velocity after both the exposure-level and the fiber-level zero points have been removed; \texttt{RV\_FLAG} records which levels a given row carries} \\
23 & \texttt{VHELIO\_CAL\_ERR\_SAME} & km\,s$^{-1}$ & \coldesc{Calibrated single-visit velocity uncertainty for same-fiber use: the pipeline \texttt{VRELERR} corrected in variance space by the same-fiber term $c$ of Section~\ref{sec:uncertainty}} \\
24 & \texttt{VHELIO\_CAL\_ERR\_CROSS} & km\,s$^{-1}$ & \coldesc{Same, using the cross-fiber term $c$} \\
25 & \texttt{RV\_FLAG} & \nodata & \coldesc{Correction level: 0 for both levels, 1 for exposure level only, 2 for uncorrected} \\
26 & \texttt{IN\_CAL} & \nodata & \coldesc{Whether the visit entered the zero-point solution (boolean)} \\
27 & \texttt{TEFF} & K & \coldesc{ASPCAP effective temperature} \\
28 & \texttt{LOGG} & \nodata & \coldesc{ASPCAP surface gravity, in dex} \\
29 & \texttt{VSINI} & km\,s$^{-1}$ & \coldesc{ASPCAP projected rotational velocity} \\
30 & \texttt{X\_H} & \nodata & \coldesc{ASPCAP metallicity, in dex} \\
31 & \texttt{FE\_H\_FLAG} & \nodata & \coldesc{Quality bitmask of the iron abundance} \\
32 & \texttt{ASPCAPFLAG} & \nodata & \coldesc{ASPCAP quality bitmask} \\
33 & \texttt{GAIAEDR3\_SOURCE\_ID} & \nodata & \coldesc{\textit{Gaia} EDR3 source identifier} \\
34 & \texttt{GAIAEDR3\_PARALLAX} & mas & \coldesc{Parallax; taken from \textit{Gaia} EDR3 for the DR17 rows and from \textit{Gaia} DR3 for the DR19 rows, so the two segments are not from the same \textit{Gaia} release} \\
35 & \texttt{EXTRATARG} & \nodata & \coldesc{Targeting-category bitmask} \\
\enddata
\end{deluxetable*}
\section{Conclusion}

APOGEE radial velocities retain a long-term systematic component that dominates on multi-year baselines. For APO repeats with $\Delta t>365$~days, the equivalent single-visit scatter of the original DRP velocities is 98.2~$\mathrm{m\,s^{-1}}$. We model this drift as the sum of an exposure-level zero point $\mu_{\rm e}$ and a fiber-level zero point $\mu_{\rm f}$, and correct each visit via $v_{\rm corr}=v-\mu_{\rm e}-\mu_{\rm f}$.

We infer both sets of zero points with an uber-calibration solver that treats each independently drifting observing stage as a calibration unit and ties units together using shared repeat-observed stars. The two correction levels are solved with the same machinery, changing only the node definition: (plate, night) for exposure-level offsets, and (fiber, observing year) for time-evolving fiber offsets (solved on exposure-corrected velocities).APO DR17 and DR19 are calibrated jointly on a common zero-point gauge, while LCO DR17, observed with a different spectrograph, is calibrated separately.

The largest gains occur at long time separations. For APO DR17, $\sigma_{\rm single}$ at $\Delta t>365$~days improves from 98.2 to 44.7~$\mathrm{m\,s^{-1}}$ after the two-level correction. For LCO DR17, $\sigma_{\rm single}$ improves from 81.3 to 77.1 and then to 64.4~$\mathrm{m\,s^{-1}}$ (original, exposure-only, exposure+fiber). For APO DR19, we recover an independent exposure-level solution, with $\sigma_{\rm single}$ improving from 118.5 to 92.8~$\mathrm{m\,s^{-1}}$, and to 72.5~$\mathrm{m\,s^{-1}}$ after including fiber zero points. At high signal-to-noise, cross-site comparisons for stars observed at both APO and LCO show improved consistency after correction.

The achievable improvement is ultimately limited by calibration-network connectivity: fewer calibration visits per unit and fewer repeat-observation links weaken the recovered zero points. This limits the fiber-level correction in DR19 (only 23.1\% of calibration stars have repeats, and the time span provides only three observing-year windows) and at LCO (a smaller calibration sample and sparser connectivity than APO DR17). Even for APO DR17, where the network is best connected, a residual long-baseline scatter of $\sim$45~$\mathrm{m\,s^{-1}}$ remains, reflecting the single-visit precision floor and/or temporal structure on timescales shorter than one observing year.

We release a per-visit, value-added radial-velocity catalog in which 3,336,141 visits carry the full two-level correction. By largely removing long-term instrumental drift, the catalog enables more reliable RV comparisons across year-long baselines, supporting RV-variability and binary searches, cluster membership analyses, and kinematic studies of stellar streams and the Galactic disk. While developed for APOGEE, the approach is broadly applicable to multi-object, fiber-fed surveys with repeat observations, and should become more effective as repeat coverage increases and calibration networks densify.

\vspace{7mm} \noindent {\bf Acknowledgments}
We acknowledge useful discussions with Prof. Cunying Xiao.
This work is supported by the National Key R\&D Program of China via 2024YFA1611901 and 2024YFA1611601, and the National Natural Science Foundation of China through the projects 12222301, and 12173007.

\bibliography{ref}
\bibliographystyle{aasjournal}

\end{document}